# Scalable alloy-based sputtering of high-conductivity $PdCoO_2$ for advanced interconnects

Takayuki Harada[1]*, Zuin Ping Lily Ang[1], Yuki Sakakibara[1], Takuro Nagai[2], Yasushi Masahiro[3]

[1] International Center for Materials Nanoarchitectonics (MANA), National Institute for Materials Science; Tsukuba, Ibaraki 305-0044, Japan.

[2] Electron Microscopy Unit, National Institute for Materials Science; Tsukuba, Ibaraki 305-0044, Japan.

[3] TANAKA Kikinzoku Kogyo K.K.; Tokyo 103-0025, Japan.

*Corresponding author. Email: HARADA.Takayuki@nims.go.jp

**Abstract**

As integrated circuits continue to scale down, the search for new metals is becoming increasingly important due to the rising resistivity of traditional copper-based interconnects. A layered oxide $PdCoO_2$ is one of the candidate materials for interconnects, having bulk *ab*-plane conductivity exceeding that of elemental Al. Despite its potential, wafer-scale vacuum deposition of $PdCoO_2$, crucial for interconnect applications, has not yet been reported. In this study, we succeeded in the scalable growth of c-axis oriented $PdCoO_2$ thin films via reactive sputtering from Pd-Co alloy targets. Our method paves the way to harness the unique properties of $PdCoO_2$ in semiconductor devices.

As the miniaturization of logic circuits and memory cells continues,[1] interconnect minimum line width is expected to reach the sub-10 nm range in future technology nodes.[2,3] In such narrow line widths, conventional Cu interconnects suffer from a serious increase in resistivity ($\rho$) as Cu needs high-resistivity diffusion barriers and adhesion liners that occupy a significant fraction of metallization volume.[2] To overcome this issue, alternative metals have been searched, which do not need barrier and liner layers and exhibit better scaling behaviors than Cu. There have been intensive studies on elemental metals[4] and alloys,[5,6] considering resistivity in narrow lines, reliability, thermal properties, process integrality, and sustainability.[2] Regrading resistivity scaling behavior, there is increasing interest in anisotropic conductors such as quasi-two dimensional (q-2D) MAX phases (*e.g.* $Cr_2AlC$ and $V_2AlC$), metallic delafossites (*e.g.* $PdCoO_2$ and $PtCoO_2$), and quasi-one dimensional (q-1D) conductors (*e.g.* CoSn and $YCo_3B_2$), which could outperform conventional Cu/liner stacks when being scaled down.[2,3,7]

Metallic delafossites such as $PdCoO_2$ (Fig. 1a) and $PtCoO_2$ are known to exhibit the lowest resistivity among q-2D conductors,[8-11] which makes them promising candidates for alternative interconnect materials.[2,3] The room-temperature resistivity in *ab*-plane directions is $\rho_{ab}$ = 2.6 μΩcm for $PdCoO_2$ and $\rho_{ab}$ = 2.1 μΩcm for $PtCoO_2$, as reported for bulk single crystals.[9] The crystal structure of $PdCoO_2$ consists of alternating $Pd^+$ and $[CoO_2]^-$ layers (Fig. 1a). The electrical transport is dominantly contributed by the conductive sheets of two-dimensional triangular lattices of $Pd^+$ ions.[9] Having remarkable q-2D conductivity and a simple Fermi surface with a hexagonal cylinder-like shape,[8-10] $PdCoO_2$ and $PtCoO_2$ have attracted significant interest as a model system to explore electrical transport phenomena.[9,10] The q-2D electrons in these systems, possessing long mean free path $l \simeq$ 20 μm for $PdCoO_2$ at cryogenic temperatures,[12] exhibit various exotic electrical conduction, including hydrodynamic motion of electrons,[13] long phase coherence length,[14] and directional ballistic transport reflecting the facetted Fermi surface.[15] Recently, these compounds are also proposed as promising materials for various applications in electronics, catalysis, and optics.[2,3,11,16-18] For interconnect applications,[2,3] we could exploit the q-2D high conductivity,[9] chemical/thermal stability,[17] and rigid oxide framework with interlayer ionic bonds (Fig. 1a) of metallic delafossites.

To incorporate $PdCoO_2$ into practical applications, scalable thin film growth techniques are crucial.[11] Initial report of $PdCoO_2$ thin films employed solid-phase reaction of amorphous precursors prepared by sputtering.[19] These thin films tend to have resistivity of $\rho \sim 10^{-4}$ $\Omega$cm,[19] several tens of times higher than the bulk *ab*-plane resistivity, possibly due to domain boundaries. Recently, c-axis oriented $PdCoO_2$ thin films with lower resistivity have been grown by pulsed laser deposition (PLD),[16,20] molecular beam epitaxy (MBE),[21] solution-based process,[22] and sputtering using sintered $PdCoO_2$ targets.[23]

Here, we demonstrate that high-quality $PdCoO_2$ thin films can be grown from a Pd-Co alloy target by reactive sputtering. Using the Pd-Co alloy target has a significant advantage compared with using the sintered $PdCoO_2$ target reported in our previous work,[23] not only in easiness of target preparation but also in thin-film quality. The typical growth conditions include a substrate temperature of $T_{growth}$ = 700°C and total gas (Ar:$O_2$ = 4:14) pressure of $P_{total}$ = 0.15–0.2 Pa. The $P_{total}$ is orders of magnitude lower than that of the previous sputtering growth using sintered $PdCoO_2$ targets[23] and PLD[16,20] (typically 13–230 Pa). The low $P_{total}$ enables the higher growth rate up to 1 nm/min depending on the RF power (Fig. S1). The optimal growth rate used to grow the $PdCoO_2/Al_2O_3$ wafer in Fig. 1 is 0.6 nm/min, which is higher than the typical growth rate by PLD ~ 0.2 nm/min (ref.[16,20]).

Figure 1(b) shows the typical $PdCoO_2$ thin-film surface marked by triangular flat domains, which is indicative of the trigonal crystal structure inherent to $PdCoO_2$. These triangular domains are consistent with the report using PLD,[16] MBE,[24] and solution-based processes.[22] The step height at the triangular domain edge predominantly corresponds to 1/3 unit cell height (u.c.), *i.e.*, a single set of Pd and $CoO_2$ layers (Fig. 1(b), bottom). The existence of additional 1/6 u.c. steps, as evidenced by terraces between grey lines in Fig.1(b), bottom, suggests a mixed termination of the surface. The x-ray diffraction (XRD) pattern and a photograph of a 2-inch $PdCoO_2/Al_2O_3$ wafer is shown in Fig. 1(c).

The XRD peaks correspond to (000$l$) reflection of PdCoO$_2$, showing that the thin-film is c-axis-oriented and the density of impurity phases is as low as the noise level of the XRD measurement. Continuous Laue oscillations over the wide range indicate the high crystalline quality and the atomic terrace flatness of the PdCoO$_2$/Al$_2$O$_3$ wafer. The XRD $\phi$ scan around PdCoO$_2$ (01$\bar{1}$2) reflection (not shown) confirmed that the PdCoO$_2$ thin films are twinned with 180°-rotated domains as observed for other growth methods.[11]

Figures 1(d) and 1(e) show cross-sectional high-angle annular dark-field scanning transmission electron microscopy (HAADF-STEM) images of the PdCoO$_2$ thin films. As shown in the crystal model of PdCoO$_2$ in Fig. 1(d), the brighter and the darker dots in the HAADF-STEM image correspond to Pd and Co atoms, respectively. The HAADF-STEM image clearly shows the characteristic layered crystal structure of PdCoO$_2$. The top and bottom surfaces of the PdCoO$_2$ thin film are indicated by arrows in Figs. 1(d) and 1(e). The alternation of Pd$^+$ and [CoO$_2$]$^-$ layers that start with the initial [CoO$_2$]$^-$ layer on the Al$_2$O$_3$ substrate persists up to the top surface of the PdCoO$_2$ thin film, indicating the high crystallinity of the entire thin film and the suppressed segregation (lower magnification images are available in Figs. S10 and S11). According to the HAADF-STEM image, the $a$-axis lattice constant of PdCoO$_2$ is 2.85 Å for the Pd layer closest to the bottom surface, whereas it is 2.84 Å for that closest to the top surface. These values are in better agreement with the bulk lattice constant of 2.83 Å than that for the corresponding length of the Al$_2$O$_3$ substrate (2.75 Å), which indicates that the PdCoO$_2$ thin film is fully relaxed from the initial growth layer.

Figure 1(f) shows the temperature-dependent resistivity of the PdCoO$_2$ thin films. The room-temperature resistivity ($\rho_{ab}^{film}$(300 K)) of the PdCoO$_2$ film ($d$ = 161.3 nm) was $\rho_{ab}^{film}$(300 K) = 4.71 μΩcm. The resistivity was reduced to $\rho_{ab}^{film}$(300 K) = 3.49 μΩcm by post annealing at 800°C for 12 h, approaching the resistivity of the bulk single crystal of $\rho_{ab}^{bulk}$(300 K) = 2.6 μΩcm (ref.[9]). By cooling the annealed thin film down to 2 K, the resistivity decreased to $\rho_{ab}^{film}$(2 K) = 0.457 μΩcm. The 27.8-nm-thick PdCoO$_2$/Al$_2$O$_3$ wafer shown in Fig. 1(a) has a room-temperature resistivity of $\rho_{ab}^{film}$(300 K) = 4.07 µΩcm. The residual resistivity ratio of the PdCoO$_2$/Al$_2$O$_3$ wafer is calculated to be $\rho_{ab}^{film}$(300 K)/$\rho_{ab}^{film}$(2 K) = 7.9. The ab-plane mean free path $l_{ab}^{film}$ of the electrons is calculated using the formulas $l_{ab}^{film} = \hbar k_F \mu_H/e$ and $\mu_H = R_H/\rho_{ab}^{film}$, where $\hbar$ is the reduced Planck constant, $k_F$ is the Fermi wavenumber, $\mu_H$ is the Hall mobility, $e$ is the elementary charge, $R_H$ is the Hall coefficient. Using the $R_H$ determined by Hall measurements (Fig. S2) and the Fermi-surface averaged $k_F = 0.96$ Å$^{-1}$ (ref.[9]), the ab-plane mean free path of the electrons at 2 K and 300 K are determined to be $l_{ab}^{film}$ (2 K) ≃ 290 nm and $l_{ab}^{film}$(300 K) ≃ 22 nm, respectively.

The quality of the PdCoO$_2$ thin films strongly depends on the preparation method of the sputtering target. We tested two types of targets, "mix target" and "melt target," prepared using distinct procedures. The mix target was prepared by sintering a mixture of Pd and Co powders sieved with a 106 µm mesh. The melt target was prepared by sintering powders of a PdCo alloy fabricated through the melt atomization of a Pd and Co mixture. According to the analysis by XRD and SEM-EDX (Fig. S3), the mix target is inhomogeneous in chemical composition with Pd-rich and Co-rich domains while the melt target is homogeneous solid solution.

Figures 2(a) and 2(b) compare the XRD patterns of PdCoO$_2$ thin films grown under the same conditions using the melt target and the mix target, respectively. The XRD pattern of the thin film fabricated from the melt target (Fig. 2(a)) shows sharp (000$l$) peaks of PdCoO$_2$ without any impurity phase peaks. In contrast, the PdCoO$_2$ thin films fabricated from the mix target shows various impurity phases (highlighted in blue in Fig. 2(b)) in addition to the PdCoO$_2$ (000$l$) peaks. The impurity peaks correspond to the Co$_3$O$_4$, PdO, and Pd$_2$O phases. Although the Pd$_2$O phase is rare in the bulk form, it stabilizes on the Al$_2$O$_3$ surfaces.[23,25] The full width at half maximum (FWHM) of the XRD rocking curves for the PdCoO$_2$ (0006) peaks are 0.071° for the thin film grown from the melt target (Fig. 2(a), inset) and 0.134° for the thin film grown from the mix target (Fig. 2(b), inset).

To examine the influence of the target preparation method in the wide range of growth parameter space, we grew PdCoO$_2$ thin films using the mix and melt targets under different conditions and characterized them using XRD. We employed the normalized sum of the impurity peak heights, $I_{imp}^{norm} = \sum I_{imp}/I_{PdCoO2}$, as a measure of the impurity density in the PdCoO$_2$ thin film. Here, $I_{PdCoO2}$ represents the XRD peak intensity of the PdCoO$_2$ (0006) reflection at around 30.2°, and $\sum I_{imp}$ represents the summed intensity of the following XRD peaks: PdO (101) at around 33.8°, Co$_3$O$_4$ (222) at around 38.4°, Pd (111) at around 40.1°, Co$_3$O$_4$ (511) at around 59.3°, Pd$_2$O (220) at around 61.2°, Co$_3$O$_4$ (440) at around 65.2°, and Pd (220) at around 68.1°. The $I_{imp}^{norm}$ is mapped in Figs. 2(c) and 2(d) for thin films grown with melt and mix targets, respectively. As shown in Fig. 2(c), using the melt target, the $I_{imp}^{norm}$ is minimized around the substrate temperature of 700°C and the gas pressure of 0.15 Pa. No impurity peaks were observed under these conditions, and $\sum I_{imp}$ corresponds to the noise level of the XRD measurement (Fig. 2(a)). In contrast, $I_{imp}^{norm}$ for the mix target in Fig. 2(d) is higher than that for the melt target. Thin films without impurity XRD peaks could not be grown using the mix target under all 20 conditions indicated by red crosses in Fig. 2(d). The XRD pattern corresponds to each growth condition is available in the supplemental online material.

These results suggest that the microscopic homogeneity of the melt target is advantageous for growing the delafossite phase. The crystal structure of PdCoO$_2$ contains Pd in the rare 1+ valence state and Co in the 3+ valence state. Since the typical ionic valence of Pd is 2+, stabilizing this rare Pd$^+$ state is crucial for the growth of PdCoO$_2$. In an oxygen atmosphere, Pd atoms individually present as either the PdO phase or the reduced Pd phase, while Co atoms form CoO (Co$^{2+}$) or Co$_3$O$_4$ (Co$^{2+}$ and Co$^{3+}$). The simultaneous presence of Pd and Co in a 1:1 ratio is essential during crystal growth to stabilize the delafossite phase with rare Pd$^+$ and Co$^{3+}$ states. Our results indicate that using a melt target, shown in Fig. S3(d), with appropriate local stoichiometry is effective in minimizing impurities in the sputtered PdCoO$_2$ thin films.

In Fig. 3, we compare the resistivity of the PdCoO$_2$ thin films and various high-conductivity metal thin films. The in-plane resistivity of the PdCoO$_2$ thin films ($\rho_{ab}^{film}$(300 K), red squares) is lower than that of epitaxial alloys and elemental metals except Ag and Cu for $d$ = 13.8 nm, 27.8 nm, and 161.3 nm, indicating the potential of PdCoO$_2$ as an interconnect material. However, the $\rho_{ab}^{film}$(300 K) increases as decreasing thickness and becomes higher than various epitaxial elemental metals for $d$ = 8.7 nm. We compare this behavior with the theoretical calculation (red broken line in Fig. 3).[7] According to the approximate Fuchs-Sondheimer (F-S) and Mayadas-Shatzkes (M-S) models,[26] the resistivity of a polycrystalline thin film depends on the thickness $d$ as,

$$\rho_{calc}^{poly} = \rho_0 + \rho_0 l_i \frac{3(1-p)}{8d} + \rho_0 l_i \frac{3R}{2D(1-R)} \quad (1)$$

where $\rho_0$ is the bulk resistivity, $l_i$ is the intrinsic electron mean free path, $p$ is the surface specularity, $R$ is the grain boundary reflectivity, and $D$ is the average grain diameter. Kumar *et al*. showed that for a single crystalline ($D \to \infty$) anisotropic conductor under the diffuse surface scattering limit ($p$ = 0), the eq. (1) can be rewritten as,[7]

$$\rho_{calc}^{single} = \rho_0 + r_{film} \frac{3}{8d} \quad (2).$$

Here, $\rho_{calc}^{single}$ is the resistivity of a single-crystal, $r_{film}$ is the scaling coefficient along the most conductive direction reflecting the Fermi surface anisotropy. According to their first-principles calculation, the $r_{film}$ = 1.66 × 10$^{-16}$ Ωm$^2$ for PdCoO$_2$.[7] Using $\rho_0 = \rho_{ab}^{bulk}$(300 K) = 2.6 µΩcm (ref.[9]), resistivity of a single crystal PdCoO$_2$ thin film $\rho_{calc}^{single}$ can be calculated as the red broken line in Fig. 3. Compared with the calculated resistivity $\rho_{calc}^{single}$ for PdCoO$_2$ single-crystal thin films, the resistivity of our thin films $\rho_{ab}^{film}$(300 K) is considerably higher and exhibits stronger dependence on $d$.

Several factors may contribute to higher $\rho_{ab}^{film}$(300 K) relative to $\rho_{calc}^{single}$. Since the calculations are based on an ideal single crystal while our thin films are twinned, electron scattering at twin domain boundaries is an obvious contributing factor.[27] This corresponds to the third term in eq. (1). The intrinsic mean free path in *ab*-plane can be calculated as, $l_i = \frac{\hbar k_F}{e} \times \frac{1}{eN^{bulk}\rho_{ab}^{bulk}} \simeq 62$ nm, using the bulk carrier concentration $N^{bulk} = 2.45 \times 10^{22}$ cm$^{-3}$. To quantitatively assess the impact of twin domain boundary scattering, we need to determine $D$ and $R$ in eq. (1), which requires microscopic understanding of twin domain structures. In addition to twin boundaries, step edges in $Al_2O_3$ substrates could affect the lateral connectivity of conducting layers[28] and result in an increase in resistivity in thinner films. Compared with these defects, epitaxial strain at $PdCoO_2/Al_2O_3$ likely have a minor effect on $\rho_{ab}^{film}$ as the $PdCoO_2$ thin films are fully relaxed from the initial growth layer, judging from the HAADF-STEM image in Fig. 1 (e).

In conclusion, we demonstrated the scalable growth of $PdCoO_2$ thin films by reactive sputtering using Pd-Co alloy targets. Our results indicate the importance of microscopic homogeneity in the target material to grow high-quality thin films by reactive sputtering. Although the relatively thick ($d > 13.8$ nm) $PdCoO_2$ films showed lower resistivity than most of the epitaxial elemental metals and alloys, the observed thickness dependence of resistivity was more pronounced than expected based on the anisotropic Fermi surface of $PdCoO_2$. Further studies are needed to reduce the resistivity of $PdCoO_2$ thin films and to integrate them with amorphous low $k$ insulators in integrated circuits in a manner compatible with the thermal budget of the back-end-of-line process.

**Acknowledgments:** A part of this work was supported by ARIM of MEXT (JPMXP1223NM5155), MEXT Leading Initiative for Excellent Young Researchers (JPMXS0320200047), JST PRESTO (JPMJPR20AD), and Grant-in-Aid for Scientific Research (B) from JSPS (24K01353).

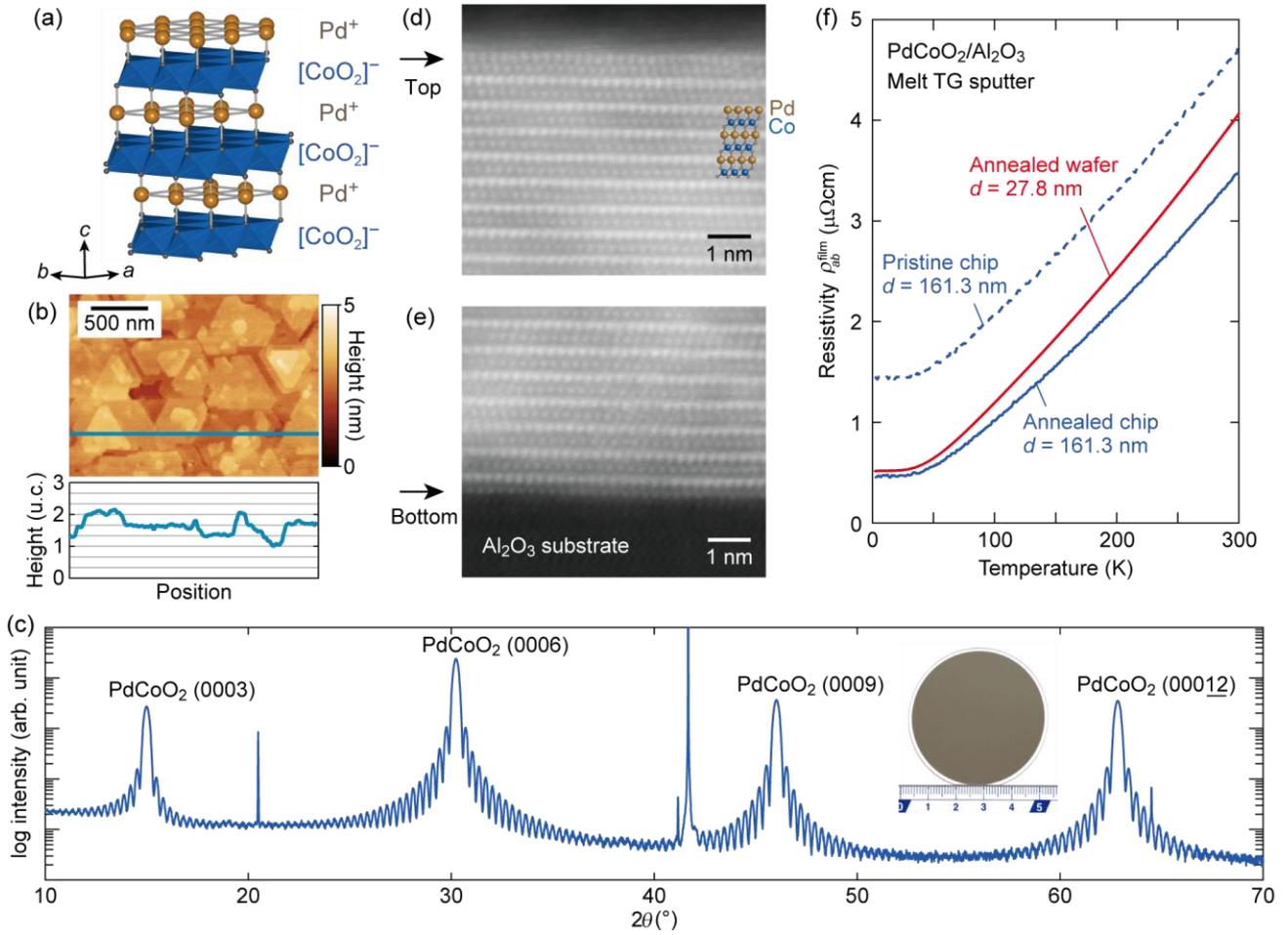

**FIG. 1.** (a) The crystal model of $PdCoO_2$ highlighting the alternating $Pd^+$ and $[CoO_2]^-$ layers. (b) Top: the surface morphology of a 2-inch $PdCoO_2$ wafer ($d$ = 27.8 nm) grown with the RF power of 100 W and post-annealed at 800°C, characterized by atomic force microscopy (AFM). Bottom: Cross-section height profile along the blue line in the AFM image. The grey horizontal lines correspond to 1/3 of the unit cell (u.c.) height along the c-axis. (c) XRD $2\theta$ -ω scan of the $PdCoO_2/Al_2O_3$ wafer. Inset: the photograph of the $PdCoO_2/Al_2O_3$ wafer. (d),(e) Cross-sectional HAADF-STEM images of a $PdCoO_2$ thin film. The top (d) and bottom surfaces (e) of the $PdCoO_2$ thin film are indicated by arrows. The crystal structure of $PdCoO_2$ is overlapped on the STEM image in (d). (f) Resistivity $\rho_{ab}^{film}$ as a function of temperature for $PdCoO_2$ thin films, which shows variations with film thickness $d$ and effect of post annealing.

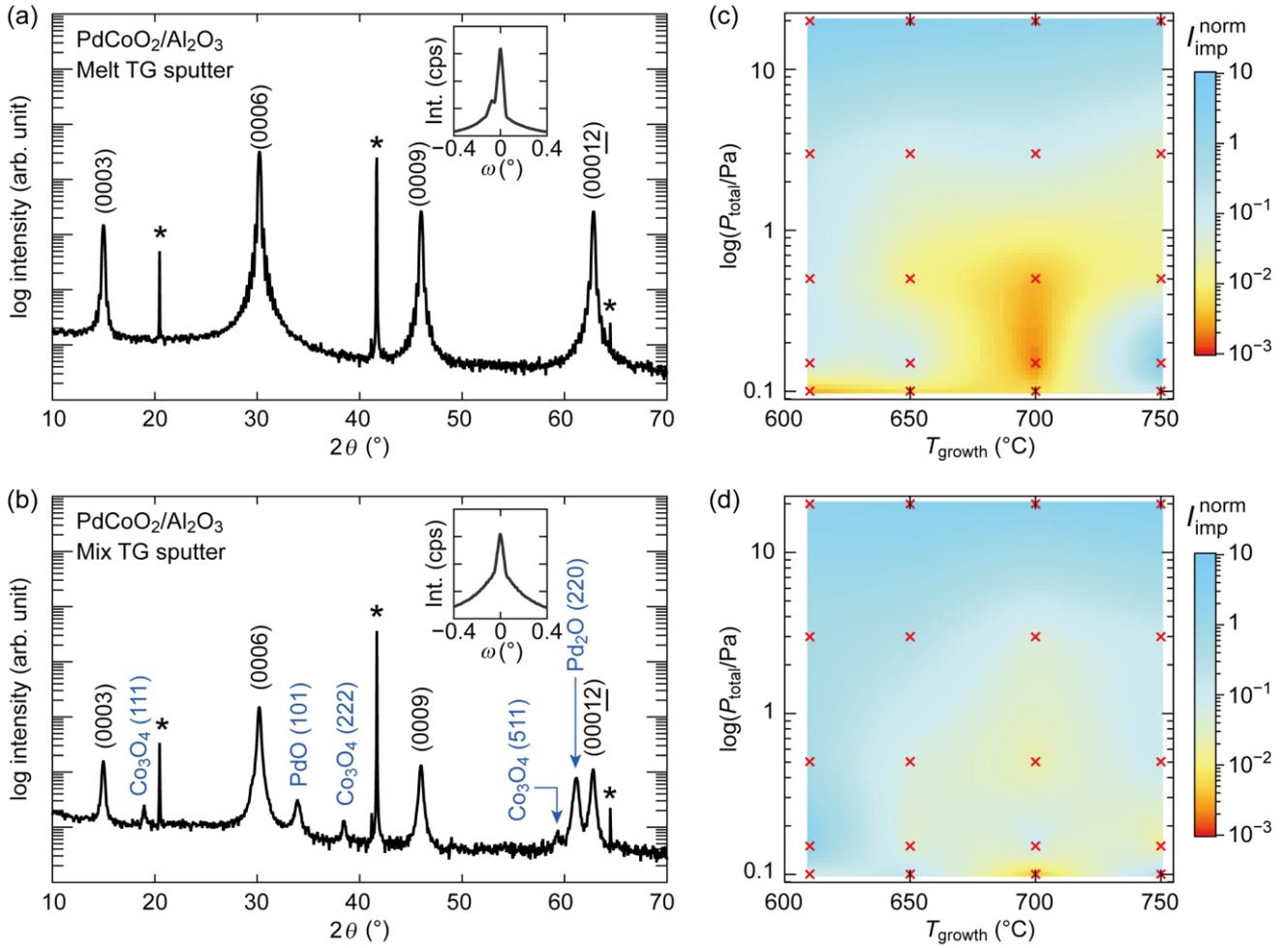

**FIG. 2.** (a) XRD $2\theta$-$\omega$ scan of a PdCoO$_2$ thin film ($d$ = 34.2 nm) grown on a sapphire chip (10×5 mm$^2$) using the melt target under optimal conditions: substrate temperature of 700°C, gas pressure of 0.15 Pa, and Ar:O$_2$ gas ratio of 4:14. (b) XRD $2\theta$-$\omega$ scan of a PdCoO$_2$ thin film grown using the mix target, under the same growth conditions as (a). The insets in (a) and (b) are the XRD rocking curves around the PdCoO$_2$ (0006) peaks, plotted in linear scale. (c) Normalized sum of impurity XRD peak intensities relative to the PdCoO$_2$ (0006) peak $I_{imp}^{norm} = \sum I_{imp}/I_{PdCoO2}$, mapped across a range of growth parameters for films grown with the melt target. The cross marks correspond to the experimental data points used to create the map. (d) Similarly plotted map for thin films fabricated with the mix target. The XRD data used to generate the maps in (c) and (d) are available in the supplemental online material.

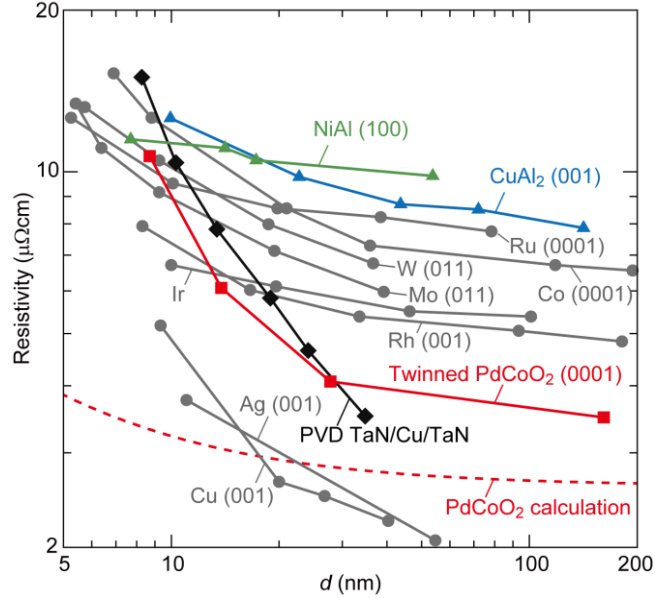

**FIG. 3.** Comparison of the resistivity of the PdCoO$_2$ thin films at 300 K ($\rho_{ab}^{film}$(300 K), red squares) with epitaxial thin films of various high-conductivity metals: elemental metals (grey circles)[4] and NiAl (green triangles)[29] at room temperature and CuAl$_2$ (blue triangles)[6] at 295 K. The surface orientation of the thin films is noted using Miller indices. The TaN/Cu/TaN (black diamonds)[2] grown by physical vapour deposition (PVD) is also plotted as a reference for the current interconnect structure. The resistivity of single-crystal PdCoO$_2$ thin films is calculated using eq. (2) and plotted in the red broken line ($\rho_{calc}^{single}$) for comparison. The numeric values of the $\rho_{ab}^{film}$ measured in this study are summarized in Table S1 in the supplemental online material.